\documentclass[11pt,aps,a4paper,eqsecnum,amsmath,amssymb
              ,nofootinbib,longbibliography
              ,notitlepage
              ]{revtex4-1}

\pdfoutput=1

\usepackage{color}
\usepackage{amsfonts}
\usepackage{graphicx}
\usepackage{amsmath}

\begin{document}

\preprint{}
\title{%
    Signatures of non-Abelian anyons in the thermodynamics\\ of an interacting fermion model
}

\author{Daniel Borcherding}
\author{Holger Frahm}
\affiliation{%
Institut f\"ur Theoretische Physik, Leibniz Universit\"at Hannover, Appelstra\ss{}e 2, 30167 Hannover, Germany}


\begin{abstract}
The contribution of anyonic degrees of freedom emerging in the non-Abelian spin sector of a one-dimensional system of interacting fermions carrying both $SU(2)$ spin and $SU(N_f)$ orbital degrees of freedom to the thermodynamic properties of the latter is studied based on the exact solution of the model.  For sufficiently small temperatures and magnetic fields the anyons appear as zero energy modes localized at the massive kink excitations [A. M. Tsvelik, Phys. Rev. Lett. \textbf{113} (2014) 066401].  From their quantum dimension they are identified as $SU(2)_{N_f}$ spin-$\frac{1}{2}$ anyons.  The density of kinks (and anyons) can be controlled by an external magnetic field leading to the formation of a collective state of these anyons described by a $Z_{N_f}$ parafermion conformal field theory for large fields.  Based on the numerical analysis of the thermodynamic Bethe ansatz equations we propose a phase diagram for the anyonic modes.
\end{abstract}

\maketitle


\section{Introduction}
The potential use of non-Abelian anyons as resources for quantum computing \cite{Kita03,NSSF08} has driven the search for physical realizations of these exotic objects.  Candidates for such systems supporting fractionalized quasi-particles with exotic statistics are the topologically ordered phases of two-dimensional matter such as fractional quantum Hall states or $p+ip$ superconductors \cite{MoRe91,ReRe99,ReGr00}.  Various one-dimensional systems have been shown to form closely related topological phases where non-Abelian anyons are realized as localized, topologically protected zero-energy modes bound to quasi-particles or defects \cite{Kitaev01,AsNa12,Tsve14a}.  The simplest of these models is the $Z_2$-invariant quantum Ising chain where the appearance of zero energy edge modes can be understood using free fermion techniques.  Signatures for these localized Majorana anyons have been found in studies of heterostructures such as semiconductor quantum wires in proximity to superconductors \cite{MZFP12,Deng.etal12}.  

Extensions of this scenario to obtain non-Abelian anyons beyond these Majorana zero modes requires dealing with interacting systems.  A possible approach is based on one-dimensional clock models, $n$-state generalizations of the Ising chain with $Z_{n>2}$ symmetry.  In the presence of chiral interactions these chains support zero energy edge modes which can be expressed in terms of $Z_n$ parafermions \cite{Fendley12,JMAF14,ZCTH15,ARFGB16,IeMM17,MPSK17}.
An alternative approach where these parafermion zero modes appear as emergent degrees of freedom starting from interacting fermions carrying both spin and orbital degrees of freedom has been proposed in Ref.~\cite{Tsve14a,Tsve14b}: 
in an integrable one-dimensional system of fermions perturbed by a marginally relevant anisotropic $SU(2)$ current-current interaction zero modes with non-integer degeneracy (i.e.\ the quantum dimension of the anyons) have been shown to reside on solitonic 'kink' excitations with finite mass.  The latter connect the topologically different ground states of the fermion model.  At temperatures well below this scale the kinks form a dilute gas.  In a real system they may get pinned to defects resulting in localized non-Abelian excitations.

\begin{figure}
  \includegraphics[width=0.9\textwidth]{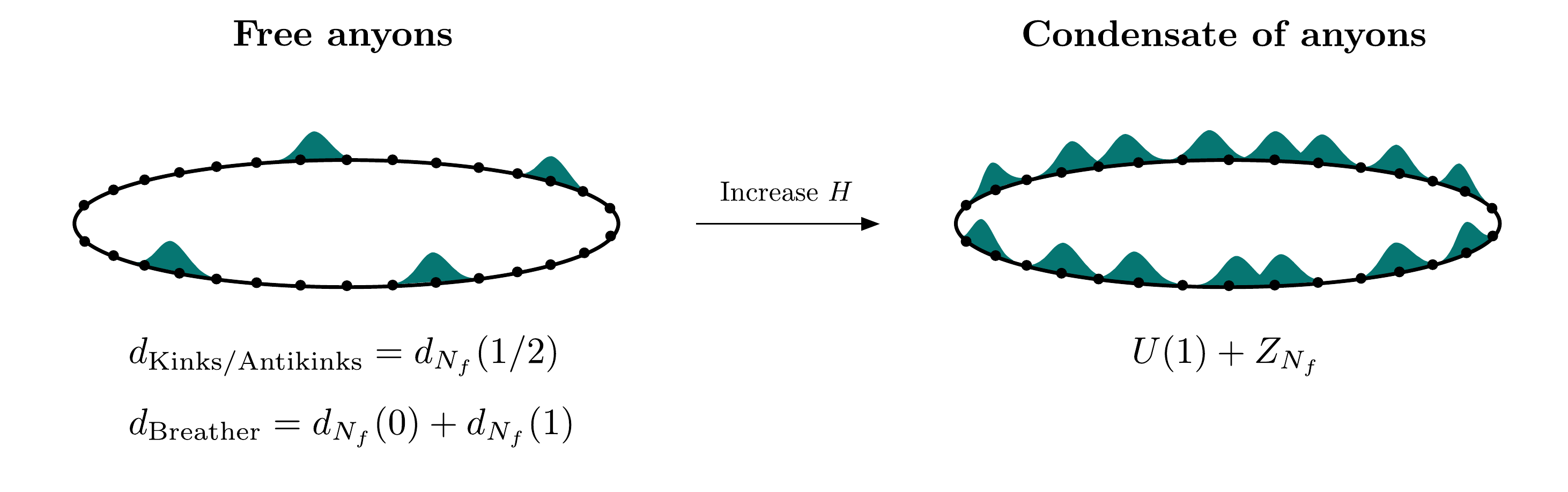}

  \caption{Illustration showing the lattice (black dots) of $N_f$-flavor fermions  and its quasi-particle excitations in the spin sector (green wave packets). For sufficiently small magnetic fields and temperatures below the energy gap of the quasi-particles
  we find a dilute gas of noninteracting anyons bound to kink, antikink, and breather excitations with their corresponding quantum dimensions (\ref{eq:qdim}). As the magnetic field is increased the kinks acquire a finite density and form a condensate.  Antikink and breathers do not contribute to the low energy spectrum which is described by a bosonic field and a non-Abelian $Z_{N_f}$ parafermion.
    \label{fig:illustration}}
\end{figure}

An additional feature of this construction is that the mass (and therefore the density) of these objects can be controlled by the application of an external magnetic field.  This allows to study in detail the transition from the low density phase of an ideal gas of massive particles with an internal degree of freedom which does not contribute to the energy to a phase with a finite density of anyons, see Figure~\ref{fig:illustration}.
This transition into a many-anyon state where interactions lift the degeneracy of the zero modes interferes with possible implementations of quantum gates for the manipulation of the anyons in a real device.  It has motivated work aimed at the understanding of the emerging collective behaviour within effective lattice models with particular short-range interactions based on the anyonic fusion rules, see e.g. Refs.~\cite{FTLT07,GATH13,Finch.etal14,BrFF16,VeJS17}, or by identification of the degrees of freedom on the boundaries between different topological phases of two-dimensional matter \cite{GATL09,GrSc09,BaSH09,Mong14}.

In this paper we study the condensation of non-Abelian anyons within the microscopic model of fermions \cite{Tsve14a} based on the exact solution of the latter.  To keep this work self-contained we recall the Bethe ansatz \cite{Tsvelik87a,JeAZ98} before we analyze the nonlinear integral equations describing the thermodynamics to characterize the excitation spectrum.  Our results are summarized in Figure~\ref{fig:phasediag} where the different manifestions of the anyonic degrees freedom in the properties of the underlying model are assembled as function of magnetic field and temperature.

\section{Interacting fermions with zero energy modes}
Following Ref.~\cite{Tsve14a,JKLRT17} we consider a model of spin-$\frac12$ fermions carrying an orbital (or flavour) degree of freedom.
In the absence of interactions the fermion fields can be expressed in terms of the slowly varying right- and left-moving modes $R_{\alpha k}$ and $L_{\alpha k}$ (with spin index $\alpha=\uparrow,\downarrow$ and orbital quantum number $k=1,\ldots,N_f$) in the vicinity of the Fermi points.  For weak couplings preserving the $U(1)$ charge, $SU(2)$ spin, and $SU(N_f)$ orbital symmetry separately one can use conformal embedding to split the Hamiltonian into a sum of three commuting parts describing the fractionalized degrees of freedom in the collective states of the interacting fermion system \cite{KnZa84,AfLu91}: the resulting $U(1)$ charge sector is described by a Gaussian model while the non-Abelian spin and orbital degrees of freedom are given by critical $SU(2)_{N_f}$ and $SU(N_f)_2$ Wess-Zumino-Novikov-Witten (WZNW) models, respectively, each perturbed by current-current interactions.  
By choosing a fermionic density in the microscopic model to be commensurate with the underlying lattice (for a possible realization with an ultracold gas of fermionic atoms in a periodic trap potential see e.g.\ Ref.~\cite{BCLMT15}) the system becomes a Mott insulator, i.e. the charge sector acquires a gap. 
Similarly, the coupling constants in the spin and orbital sectors can be tuned such that the perturbations are marginally relevant.  As a result the corresponding excitations are massive and display non-Abelian statistics \cite{Tsvelik87a}.  
Specifically, we shall consider the scenario where the resulting energy scale corresponding to the $SU(2)_{N_f}$ spin degrees of freedom is the lowest one.  In this situation the low energy behaviour of the system can be described based on the Hamiltonian density
\begin{equation}
\label{NfModel}
    \mathcal{H} = \frac{2\pi}{N_f+2} \left(:J^a J^a: + :\bar{J}^a \bar{J}^a:\right) + \lambda_\parallel J^z\bar{J}^z + \frac12 \lambda_\perp \left( J^+ \bar{J}^- + J^- \bar{J}^+ \right)\,.
\end{equation}
We have expressed the anisotropic current-current interaction in terms of the $SU(2)_{N_f}$ Kac-Moody currents $J^a = \frac12 R^\dagger_{\alpha k} (\sigma^a)_{\alpha\beta} R_{\beta k}$ and their left-moving counterparts $\bar{J}^a$, defined similarly in terms of $L_{\alpha k}$ ($\sigma^a$, $a=x,y,z$ are Pauli matrices).   For $\lambda_\parallel>0$ this model is known to be massive, the elementary excitations are kinks and antikinks with finite mass $M_0$ (see e.g.\ Ref.~\cite{Tsvelik87a} and below).

The exact spectrum of the underlying fermion model can be obtained by means of the Bethe ansatz \cite{Tsvelik87a,JeAZ98}.  Here we concentrate on the spin sector: placing the system of $N$ fermions into a box of length $L$ with periodic boundary conditions and exposing it to a magnetic field $H$ coupled to the $z$ component of the total spin the eigenvalues of (\ref{NfModel}) are \cite{PoWi83,FaRe86,Tsve87,Wiegmann85}
	\begin{equation}
		\label{energy}
		E=\sum_{\alpha=1}^{M}\left(\sum_{\tau=\pm 1}\frac{\tau}{2\rm{i}}\ln\left(\frac{\sinh(\frac{\pi}{2p_0}(\lambda_\alpha+\tau/g-N_f\rm{i}))}{\sinh(\frac{\pi}{2p_0}(\lambda_\alpha+\tau/g+N_f\rm{i}))} \right)+H\right) - \frac{N\,H}{2}\,,
	\end{equation}
where $g$ and $p_0$ are functions of the coupling constants $\lambda_\parallel$ and $\lambda_\perp$.
In a sector with magnetization $S^z=(N/2)-M$, the energies (\ref{energy}) are parametrized by the complex parameters $\lambda_\alpha$, $\alpha=1,\ldots,M$, solving the Bethe equations
	\begin{equation}
	\label{Betheeq}
		\prod_{\tau=\pm 1}\left(\frac{\sinh\left(\frac{\pi}{2p_0}(\lambda_\alpha+\tau/g+N_f\rm{i})\right)}{\sinh\left(\frac{\pi}{2p_0}(\lambda_\alpha+\tau/g-N_f\rm{i})\right)} \right)^{N/2}
		=\prod_{\beta=1}^{M}\frac{\sinh(\frac{\pi}{2p_0}(\lambda_\alpha-\lambda_\beta+2\rm{i}))}{\sinh(\frac{\pi}{2p_0}(\lambda_\alpha-\lambda_\beta-2\rm{i}))}\,.
	\end{equation}
The relativistic invariance of the low energy spectrum is broken by the boundary conditions used.  It can be restored by considering observables in the scaling limit $g\ll1$ and $L,N\to\infty$ such that the mass of the spin excitations is small compared to the particle density $N/L$. 

Note that the Bethe equations (\ref{Betheeq}) coincide with the ones obtained for an integrable spin $S=N_f/2$ XXZ chain of $N$ sites with staggered inhomogeneities \cite{KiRe86,KiRe87}.  Based on this observation the root configurations solving (\ref{Betheeq}) in the thermodynamic limit can be classified in terms of strings, i.e. groups consisting of $n$ Bethe roots
    \begin{equation}
    \label{string}
	    \lambda^n_{\alpha,j}=\lambda^n_\alpha + i\left(n+1-2j+\frac{p_0}{2}(1-v_{2S}v_{n})\right), \quad \quad j=1,\dots,n,
	\end{equation}
where $\lambda^n_\alpha \in \mathbb{R}$ is the center of the string and $v_n\in\{\pm 1\}$ is called its parity, $v_{2S}=v_{n=2S}$. The length and parity $(n_j,v_{n_j})$ of admissible strings depend on the parameter $p_0=N_f+1/\nu$, see Appendix~\ref{app:stringclassification}.  To simplify the discussion below we assume $\nu>2$ to be an integer.\footnote{%
This is a technical assumption which does not limit the applicability of the results given below: the properties of the model depend smoothly on the parameter $p_0$ in extended intervals around this value \cite{KiRe87,FrYF90}.}
Considering a root configuration consisting of $\nu_j$ strings of type $(n_j,v_{n_j})$ and using (\ref{string}) the Bethe equations (\ref{Betheeq}) can be rewritten in terms of the real string-centers $\lambda_\alpha^{(j)} \equiv \lambda_\alpha^{n_j}$.  In their logarithmic form they read
	\begin{equation}
		\label{centerEq}
		\frac{N}{2}\left(t_{j,N_f}(\lambda^{(j)}_\alpha+1/g)+t_{j,N_f}(\lambda^{(j)}_\alpha-1/g)\right)=2\pi I^{(j)}_\alpha + \sum_{k\geq 1}\sum_{\beta=1}^{\nu_k}\theta_{jk}(\lambda^{(j)}_\alpha-\lambda^{(k)}_\beta),
	\end{equation}
where $I^{(j)}_\alpha$ are integers (or half-integers) and we have introduced functions
\begin{equation}
	\begin{aligned}
		t_{j,N_f}(\lambda)&=\sum_{l=1}^{\min(n_j,N_f)}f(\lambda,|n_j-N_f|+2l-1,v_jv_{N_f})\\
		\theta_{jk}(\lambda)&=f(\lambda,|n_j-n_k|,v_jv_k)+f(\lambda,n_j+n_k,v_jv_k)+2\sum_{\ell=1}^{\min(n_j,n_k)-1}f(\lambda,|n_j-n_k|+2\ell,v_jv_k)\,
	\end{aligned}
\end{equation}
with
\begin{equation*}
    f(\lambda,n,v)=
		\begin{cases}
			 2\arctan\left(\tan((\frac{1+v}{4}-\frac{n}{2p_0})\pi)\tanh(\frac{\pi\lambda}{2p_0})\right) \quad \text{if }\frac{n}{p_0}\neq \text{integer} \\
			 0 \hspace*{6.8cm} \text{if }\frac{n}{p_0}= \text{integer}
		\end{cases}\,.
\end{equation*}

In the thermodynamic limit, $M,N\rightarrow \infty$ with $M/N$ fixed, the centers $\lambda^{(j)}_\alpha$ are distributed continuously with densities $\rho_j(\lambda)$ and hole densities $\rho^h_j(\lambda)$.  Within the root density formalism \cite{YaYa66b} the densities are defined through the following integral equations
	\begin{equation}
		\label{densities1}
		\rho^{(0)}_j(\lambda)=(-1)^{r(j)}\rho^h_j(\lambda)+\sum_{k\geq 1}A_{jk}\ast \rho_k(\lambda)\,.
	\end{equation}
Here $a*b$ denotes a convolution and $r(j)$ is determined by $m_{r(j)}\leq j< m_{r(j)+1}$, see Appendix~\ref{app:stringclassification}.  The bare densities $\rho^{(0)}_j(\lambda)$ and the kernels $A_{jk}(\lambda)$ of the integral operators are given by
\begin{equation}
	\begin{aligned}
		\rho^{(0)}_j(\lambda)&=\frac{1}{2}\left(a_{j,N_f}(\lambda+1/g)+a_{j,N_f}(\lambda-1/g)\right)\,, \quad
		a_{j,N_f}(\lambda)=\frac{1}{2\pi}\frac{d}{d\lambda}t_{j,N_f}(\lambda)\,,\\
		A_{jk}(\lambda)&=\frac{1}{2\pi}\frac{d}{d\lambda}\theta_{jk}(\lambda)+(-1)^{r(j)}\delta_{jk}\delta(\lambda)\,.
	\end{aligned}
\end{equation}
In terms of the solution of (\ref{densities1}) for the densities of strings the energy density $\mathcal{E}=E/N$ is obtained from (\ref{energy}) as
\begin{equation}
	\label{energy2}
	\begin{aligned}
		\mathcal{E}&=\frac{1}{N}\sum_{j\geq 1}\sum_{\alpha=1}^{\nu_j}\left(\frac{1}{2}(t_{j,N_f}(\lambda^{(j)}_{\alpha}+1/g)-t_{j,N_f}(\lambda^{(j)}_{\alpha}-1/g))+n_jH\right)-\frac{H}{2}\\
		&\stackrel{N\rightarrow \infty}{=}\sum_{j\geq 1}\int_{-\infty}^{+\infty}\text{d}\lambda\, \epsilon^{(0)}_j(\lambda)\rho_{j}(\lambda)-\frac{H}{2},
	\end{aligned}
\end{equation}
where we introduced the bare energies
	\begin{equation}
		\epsilon^{(0)}_j(\lambda)=\frac{1}{2}\left(t_{j,N_f}(\lambda + 1/g)-t_{j,N_f}(\lambda-1/g)\right)+n_jH\,.
	\end{equation}

In the present context with $p_0=N_f+1/\nu$ with integer $\nu>2$ there are $N_f+\nu$ allowed string configurations (\ref{string}).  The energy (\ref{energy2}) is minimized by a configuration where only $j_0$-strings of length $n_{j_0}=N_f$ have a finite density, for small magnetic fields they fill the entire real axis.  Inverting the kernel $A_{j_0j_0}$ in the equation for $\rho_{j_0}$ and inserting the result into the other equations of (\ref{densities1}) we end up with the following set of integral equations
	\begin{equation}
		\label{densityinteq}
	    \begin{aligned}
	    \rho_{j_0}(\lambda)&=\tilde{\rho}^{(0)}_{j_0}(\lambda)-B_{j_0j_0}\ast\rho^h_{j_0}(\lambda)-\sum_{k\neq j_0}B_{j_0k}\ast \rho_{k}(\lambda),\\
		\rho^h_{j}(\lambda)&=\tilde{\rho}^{(0)}_{j}(\lambda)-B_{jj_0}\ast\rho^h_{j_0}(\lambda)-\sum_{k\neq j_0}B_{jk}\ast \rho_{k}(\lambda), \quad j\in\{j_1\}\cup\{j_2\},
   	\end{aligned}
	\end{equation}
where we introduced the kernels $B_{jk}(\lambda)$ given in Appendix~\ref{AppendixB}.  Following \cite{KiRe87} holes in the distribution of $j_0$-strings with density $\rho_{j_0}^h(\lambda)$ are called kinks, the particle like excitations corresponding to the $\nu$ types of strings with $j\in\{j_1: N_f\le j_1<N_f+\nu\}$ are called breathers.  In addition there are $N_f-1$ auxiliary zero-energy modes ($j\in\{j_2\}=\{1,2,\ldots,N_f-1\}$), see Appendix~\ref{app:stringclassification}.  The bare densities of these modes entering the integral equations (\ref{densityinteq}) are (see Appendix~\ref{app:scalinglim} for details on taking the scaling limit  $g\ll1 $ for $\tilde{\rho}^{(0)}_j(\lambda)$)
\begin{equation}
		\label{relations1}
	\begin{aligned}
		\tilde{\rho}^{(0)}_{j_0}(\lambda)&\equiv\frac{1}{A_{j_0j_0}}\ast\rho^{(0)}_{j_0}(\lambda)\stackrel{g\ll 1}{=}\frac{M_{0}}{4}\cosh(\pi\lambda/2)\\
		\tilde{\rho}^{(0)}_{j_1}(\lambda)&\equiv\frac{A_{j_1j_0}}{A_{j_0j_0}}\ast\rho_{j_0}^{(0)}(\lambda)-\rho^{(0)}_{j_1}(\lambda)\stackrel{g\ll 1}{=}\frac{{M}_{j_1}}{4}\cosh(\pi\lambda/2)\\
		\tilde{\rho}^{(0)}_{j_2}(\lambda)&\equiv\rho^{(0)}_{j_2}(\lambda)-\frac{A_{j_2j_0}}{A_{j_0j_0}}\ast\rho_{j_0}^{(0)}(\lambda)=0\,,
	\end{aligned}
\end{equation}
with masses $M_{j_0} \equiv M_0 = 2 e^{-\frac{\pi}{2g}} $ and
\begin{equation}
	\begin{aligned}
		M_{j_1} &\equiv 
		\begin{cases}
			2M_{0}\,\sin\left((j_1-N_f+1)\frac{\pi}{2\nu}\right) & \text{if }N_f\le j_1 < N_f+\nu-1\,,\\
			M_{0} & \text{if }j_1= N_f+\nu-1.
		\end{cases}
	\end{aligned}
\end{equation}
In terms of the kink and breather densities the energy density is given by
	\begin{equation}
		\label{energy3}
		\mathcal{E}=E_0+\int_{-\infty}^{\infty}\text{d}\lambda\,\tilde{\epsilon}^{(0)}_{j_0}(\lambda)\rho^h_{j_0}(\lambda)+\sum_{j_1}\int_{-\infty}^{\infty}\text{d}\lambda\,\tilde{\epsilon}^{(0)}_{j_1}(\lambda)\rho_{j_1}(\lambda)\,.
	\end{equation}
Here the new bare energies are
\begin{equation}
    \label{dressedE_gs}
	\begin{aligned}
		\tilde{\epsilon}^{(0)}_{j_0}(\lambda)&\equiv\frac{1}{A_{j_0j_0}}\ast \epsilon^{(0)}_{j_0}(\lambda)\stackrel{g\ll 1}{=}M_{0}\cosh(\pi\lambda/2)-zH\,,\\
		\tilde{\epsilon}^{(0)}_{j_1}(\lambda)&\equiv\epsilon^{(0)}_{j_1}(\lambda)-\frac{A_{j_0j_1}}{A_{j_0j_0}}\ast \epsilon^{(0)}_{j_0}(\lambda)
		\stackrel{g\ll 1}{=}M_{j_1}\cosh(\pi\lambda/2)+zH\,\delta_{j_1,j_0-1}\,,\\
		\tilde{\epsilon}^{(0)}_{j_2}(\lambda)&\equiv\epsilon^{(0)}_{j_2}(\lambda)-\frac{A_{j_0j_2}}{A_{j_0j_0}}\ast \epsilon^{(0)}_{j_0}(\lambda)=0\,,
	\end{aligned}
\end{equation}
with $z\equiv n_{j_0}/A_{j_0j_0}(0)=\frac{1}{2}(1+N_f\nu)$.  $E_0$ is the ground state energy density
	\begin{equation}
		E_0=\int_{-\infty}^{\infty}\text{d}\lambda \tilde{\epsilon}^{(0)}_{j_0}(\lambda)\rho^{(0)}_{j_0}(\lambda) -\frac{H}{2}\,.
	\end{equation}	
Note that the breather with $j_1=N_f+\nu-1 \equiv \tilde{j}_0$ has the same mass $M_{j_0}$ as the kink.  Its coupling to the magnetic field, however, is with the opposite sign.  Therefore, following Ref.~\cite{Tsve87}, we denote this breather 'antikink' below.

\section{Thermodynamics}
\begin{figure}
  \includegraphics[width=0.6\textwidth]{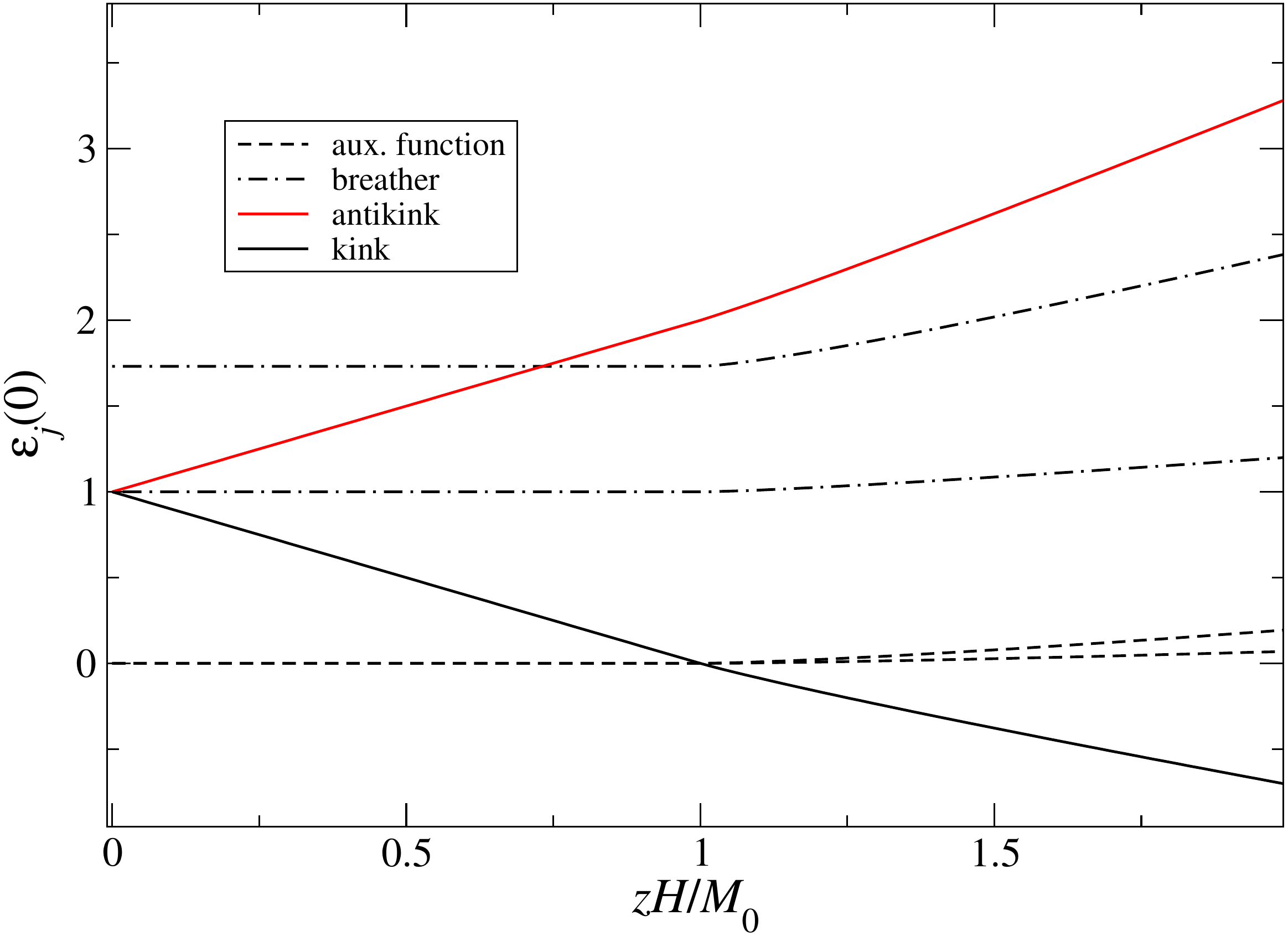}
  
  \includegraphics[width=0.6\textwidth]{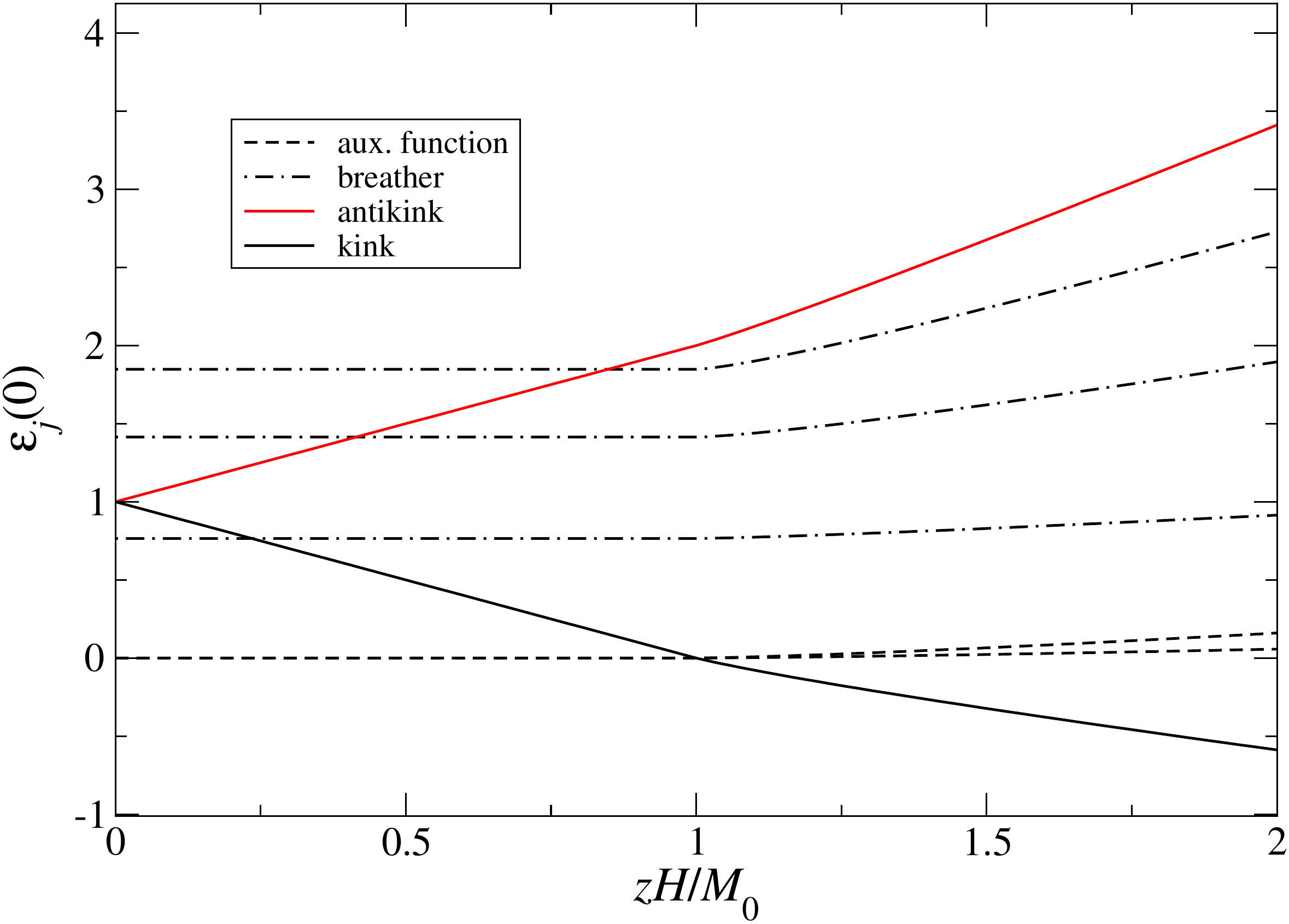}

  \caption{The zero temperature spectrum of excitations (and Fermi energy of the kinks for $zH>M_0$, respectively) $\epsilon_j(0)$ obtained from the numerical solution of (\ref{dressedenergyinteq}) as function of the magnetic field for $N_f=3$.  The number and masses of the breather modes depend on the value of $\nu$ being $3$ (top) and $4$ (bottom), respectively.  The low lying modes with energies described by the auxiliary functions are clearly separated from the spectrum of kinks, antikinks and breathers.  Their degeneracy is lifted as soon as the kink gap closes.
    \label{fig:spec}}
\end{figure}

For the physical properties of the different quasi-particles appearing in the Bethe ansatz solution of the model (\ref{NfModel}) we study its thermodynamics.  The equilibrium state at finite temperature is obtained by minimizing the free energy, $F/N=\mathcal{E}-T\mathcal{S}$, with the combinatorial entropy \cite{YaYa69}
	\begin{equation}
		\label{entropy}
		\mathcal{S}=\sum_{j\geq 1}\int_{-\infty}^{+\infty}\text{d}\lambda\,
		\left[(\rho_j+\rho^h_j)\ln(\rho_j+\rho^h_j)-\rho_j\ln\rho_j-\rho^h_j\ln\rho^h_j\right]\,.
	\end{equation}
The resulting thermodynamic Bethe ansatz (TBA) equations read
	\begin{align}
		\label{dressedeinteq}
		&T\ln(1+e^{\epsilon_k/T})=\tilde{\epsilon}^{(0)}_{k}(\lambda) +\sum_{j\geq 1} B_{jk}\ast T\ln(1+e^{-\epsilon_j/T}),
	\end{align}
where we have introduced the dressed energies $\epsilon_j(\lambda)$ through $e^{-\epsilon_j/T}=\rho_j/\rho^h_j$ for breathers and auxiliary modes $j\in\{j_1\}\cup\{j_2\}$ and $e^{-\epsilon_{j_0}/T}= \rho^h_{j_0}/\rho_{j_0}$ for kinks.  In terms of the dressed energies the free energy per particle is
\begin{equation}
    \label{freeenergy}
    \begin{aligned}
	\frac{F}{N}&\,\,=\,\,-T\sum_{j \notin\{j_2\}} \int_{-\infty}^{\infty}\text{d}\lambda\, \tilde{\rho}^{(0)}_{j}(\lambda)\ln(1+e^{-\epsilon_{j}/T})\,\\
	&\stackrel{g\ll 1}{=}-\frac{T}{4}\sum_{j \notin\{j_2\}} M_{j}\int_{-\infty}^{\infty}\text{d}\lambda\, \cosh(\pi\lambda/2)\ln(1+e^{-\epsilon_{j}/T})\,.
	\end{aligned}
\end{equation}
Finally, the corresponding integral equations determining the dressed energies $\epsilon_{j_0}$, $\epsilon_{\tilde{j}_0}$ of the kink and antikink, $\epsilon_{j_1}$ of the breathers ($j_1\neq \tilde{j}_0$), and $\epsilon_{j_2}$ of the auxiliary modes are
\begin{equation}
	\label{dressedenergyinteq}
	\begin{aligned}
		\epsilon_{j_0}(\lambda)&=M_{0}\cosh(\pi\lambda/2)-zH+\sum_{k\geq 1} (B_{kj_0}-\delta_{kj_0})\ast T\ln(1+e^{-\epsilon_k/T})\\
		\epsilon_{\tilde{j}_0}(\lambda)&=M_{0}\cosh(\pi\lambda/2)+zH+\sum_{k\geq 1} (B_{kj_0}-\delta_{k\tilde{j}_0})\ast T\ln(1+e^{-\epsilon_k/T})\\
		\epsilon_{j_1}(\lambda)&=M_{j_1}\cosh(\pi\lambda/2)+\sum_{k\geq 1} (B_{kj_1}-\delta_{k\in\{j_1\}})\ast T\ln(1+e^{-\epsilon_k/T})\\
		\epsilon_{j_2}(\lambda)&=\sum_{k\geq 1} (B_{kj_2}-\delta_{k\in\{j_2\}})\ast T\ln(1+e^{-\epsilon_k/T}).
	\end{aligned}
\end{equation}
Solving these equations we obtain the spectrum of the model (\ref{NfModel}) for a given temperature and magnetic field, see Figure~\ref{fig:spec} for $T=0$.
For magnetic fields $zH\lesssim M_{0}$ the kinks, antikinks and breathers are gapped.  As $H$ is increased the kink gap closes and they condense into a phase where they form a collective state with finite density.

We can now discuss the low-temperature behaviour of the free energy as function of the magnetic field.  

\subsection{$T\ll \min(M_{0}-zH, M_{N_f})$, $zH < M_{0}$}
For magnetic fields $zH<M_{0}$ we consider temperatures below the gaps of the kink and the lowest energy breather ($j_1=N_f$), i.e.\ $T\ll \min(M_{0}-zH,M_{N_f})$.  In this regime the nonlinear integral equations (\ref{dressedenergyinteq}) can be solved iteratively \cite{Tsve14a}: the energies of kinks, antikinks, and breathers, $\epsilon_{j_0}, \, \epsilon_{\tilde{j}_0}$ and $\epsilon_{j_1}$ are well described by their first order approximation while those of the auxiliary modes can be replaced by the asymptotic solution for $|\lambda|\rightarrow \infty$
	\begin{equation}
		1+e^{\epsilon_{j_2}/T}=\left(\frac{\sin(\frac{\pi(j_2+1)}{N_f+2})}{\sin(\frac{\pi}{N_f+2})}\right)^2\,.
	\end{equation}
For the other modes we obtain ($Q=2\cos(\pi/(N_f+2))$)
\begin{equation}
	\begin{aligned}
		\epsilon_{j_0}(\lambda)&=M_{0}\cosh(\pi\lambda/2)-zH-T\ln Q\\
		\epsilon_{\tilde{j}_0}(\lambda)&=M_{0}\cosh(\pi\lambda/2)+zH-T\ln Q\\
		\epsilon_{j_1}(\lambda)&={M}_{j_1}\cosh(\pi\lambda/2)-T\ln Q^2
	\end{aligned}
\end{equation}
resulting in the free energy
\begin{equation}
    \label{eq:Fideal}
	\begin{aligned}
		\frac{F}{N}=&-TQ\int\frac{\text{d}p}{2\pi}e^{-(M_{0}-zH)/T-p^2/2M_{0}T}-TQ\int\frac{\text{d}p}{2\pi}e^{-(M_{0}+zH)/T-p^2/2M_{0}T}\\
		&-TQ^2\sum_{j\in \{j_1\}\setminus \{\tilde{j}_0\}}\int\frac{\text{d}p}{2\pi}e^{-M_{j}/T-p^2/2M_{j}T}\,.
	\end{aligned}
\end{equation}
As observed in Ref.~\cite{Tsve14a} each of the terms appearing in this expression is the free energy of an ideal gas of particles with the corresponding mass carrying an internal degree of freedom with non-integer 'quantum dimension' $Q$ for the kinks (and antikinks) and $Q^2$ for the breathers.  Their densities
\begin{equation}
	\begin{aligned}
		&n_{j_0}=Q\sqrt{\frac{M_{0}T}{2\pi}}e^{-(M_{0}-zH)/T}, \quad n_{\tilde{j}_0}=Q\sqrt{\frac{M_{0}T}{2\pi}}e^{-(M_{0}+zH)/T}\\
		&n_{j_1}=Q^2\sqrt{\frac{M_{j_1}T}{2\pi}}e^{-M_{j_1}/T}, \quad j_1 \neq \tilde{j}_0
	\end{aligned}
\end{equation}
can be controlled by variation of the temperature and the magnetic field, which acts as a chemical potential $\pm zH$ for kinks and antikinks, respectively.

For $zH\lesssim M_{0}$ the dominant contribution to $F$ is that of the kinks.  Their degeneracy $Q$ coincides with the quantum dimension of anyons satisfying $SU(2)_{N_f}$ fusion rules with spin $j=1/2$ \cite{Kita06,Bonderson07}\,\footnote{%
These include Ising anyons for $N_f=2$ and, due to the automorphism $j \to N_f/2-j$, Fibonacci anyons for $N_f=3$.}
\begin{equation}
    \label{eq:qdim}
  d_{N_f}(j) = \sin\left(\frac{\pi(2j+1)}{N_f+2}\right) / \sin\left(\frac{\pi}{N_f+2}\right)\,.
\end{equation}
This has been interpreted as a signature for the presence of anyonic zero modes bound to the kinks \cite{Tsve14a}.  As the field is reduced the gap of the kinks increases and for $\nu>3$, the lowest breather, $j_1=N_f$, dominates the free energy when $zH\lesssim M_{0}-M_{N_f}$.  Its degeneracy $Q^2$ can be understood as a consequence of the breather being a bound state of a kink and an antikink, each contributing a factor $Q=d_{N_f}(\frac12)$: from the fusion rule for $SU(2)_{N_f}$ spin-$\frac12$ anyons, $\frac12\times\frac12=0+1$, we obtain the degeneracy of this bound state to be $Q^2 = d_{N_f}(0)+d_{N_f}(1)$.

\subsection{$T\ll zH - M_{0}$, $zH \gg M_{0}$}
Following \cite{KiRe87b} we observe that in this regime the dressed energies and densities can be related as
\begin{equation}
		\label{derelations}
	\begin{aligned}
		\rho_j(\lambda)&=(-1)^{\delta_{j\in\{j_2\}}}\frac{1}{2\pi}\frac{\text{d}\epsilon_j(\lambda)}{\text{d}\lambda} f\left(\frac{\epsilon_j(\lambda)}{T}\right),\\
		\rho^h_j(\lambda)&=(-1)^{\delta_{j\in\{j_2\}}}\frac{1}{2\pi}\frac{\text{d}\epsilon_j(\lambda)}{\text{d}\lambda} \left(1-f\left(\frac{\epsilon_j(\lambda)}{T}\right)\right)\,,
	\end{aligned}
\end{equation}
for $\lambda> \lambda_\delta$ with $\exp(\pi\lambda_\delta/2)\gg1$, where $f(\epsilon)= (1+e^\epsilon)^{-1}$ is the Fermi function.  Inserting this into (\ref{entropy}) we get ($\phi_j=\epsilon_j/T$)
\begin{equation}
	\begin{aligned}
		\label{entropy2}
		&\mathcal{S}=\sum_{j}\mathcal{S}_j(\lambda_\delta)-\frac{T}{\pi}\sum_{j}(-1)^{\delta_{j\in\{j_2\}}} \int_{\phi_j(\lambda_\delta)}^{\phi_j(\infty)}\text{d}\phi_j\, \left[f(\phi_j)\ln f(\phi_j)+(1-f(\phi_j))\ln(1-f(\phi_j))\right]\,,\\
		&\mathcal{S}_j(\lambda_\delta)\equiv \int_{-\lambda_\delta}^{\lambda_\delta}\text{d}\lambda\, \left[(\rho_j+\rho^h_j)\ln(\rho_j+\rho^h_j)-\rho_j\ln\rho_j-\rho^h_j\ln\rho^h_j\right]\,.
	\end{aligned}
\end{equation}
The integrals over $\phi_j$ can be performed giving
	\begin{align}
		\label{entropyassymp}
		&\mathcal{S}=\sum_{j}\mathcal{S}_j(\lambda_\delta)-\frac{2T}{\pi}\sum_{j}(-1)^{\delta_{j\in\{j_2\}}}[L(f(\phi_j(\infty))-L(f(\phi_j(\lambda_\delta)))].
	\end{align}
in terms of  the Rogers dilogarithm $L(x)$
\begin{equation}
    L(x)=-\frac{1}{2}\int_{0}^{x}\text{d}y\, \left(\frac{\ln y}{1-y}+\frac{\ln (1-y)}{y}\right)\,.
\end{equation}
In  the regime considered here, i.e.\ $T\ll zH-M_0$ and $\log(zH/M_0)>\lambda_\delta\gg1$, we conclude from (\ref{dressedenergyinteq}) that
\begin{equation}
	    \label{ccondition}
	\begin{aligned}
		&f(\phi_{j_0}(\lambda_\delta))=1,\quad f(\phi_{j_1}(\lambda_\delta))=0, \quad f(\phi_{j_2}(\lambda_\delta))=0\\
		&f(\phi_{j_0}(\infty))=0, \quad f(\phi_{j_1}(\infty))=0, \quad f(\phi_{j_2}(\infty))=\left(\frac{\sin(\frac{\pi}{N_f+2})}{\sin(\frac{\pi(j_2+1)}{N_f+2})}\right)^2\,,
	\end{aligned}
\end{equation}
and therefore
\begin{equation}
		\rho_{j_0}(\lambda)=0, \quad \rho^h_{j_1}(\lambda)=0, \quad \rho^h_{j_2}(\lambda)=0\,,
		\quad \mathrm{for~} |\lambda|<\lambda_\delta\,
\end{equation}
giving $\mathcal{S}_j(\lambda_\delta)=0$ for all $j$.  Finally, using the identity
	\begin{equation*}
		\sum_{k=2}^{n-2}L\left(\frac{\sin^2(\pi/n)}{\sin^2(\pi k/n)}\right)=\frac{2(n-3)}{n}L(1), \quad L(1)=\frac{\pi^2}{6},\quad n\geq 3
	\end{equation*}
we find
	\begin{equation}
	    \label{Ent_CFT}
		\mathcal{S}=\frac{\pi}{3}\left(1+2\frac{N_f-1}{N_f+2}\right)\,T
		    + O(T^2)
		 = \frac{\pi}{3}\,\frac{3N_f}{N_f+2}\,T + O(T^2)\,
	\end{equation}
for the universal low-$T$ asymptotics of the entropy in the phase with  finite kink density.  The low energy excitations near the Fermi points $\epsilon_{j_0}(\pm\Lambda)=0$ of the kink dispersion propagate with velocity $v_{\text{kink}} = \left. (\partial_\lambda \epsilon_{j_0}) / (2\pi\rho_{j_0}^h)\right|_{\Lambda} \to 1$ for magnetic fields $H>H_\delta$ such that $\Lambda(H_\delta)>\lambda_\delta$.  Hence the conformal field theory (CFT) describing the collective low energy modes is the $SU(2)$ WZNW model at level $N_f$ or, by conformal embedding \cite{FrMSbook96,Tsve14a,JKLRT17}, a product of a free $U(1)$ boson and a $Z_{N_f}$-parafermion coset $SU(2)_{N_f}/U(1)$ contributing $c=1$ and
\begin{equation}
    \label{cZn}
	c(Z_{N_f})=2\frac{N_f-1}{N_f+2}\,
\end{equation}
to the central charge, respectively.
The latter is the critical theory for the ($N_f+1$)-state restricted solid-on-solid (RSOS) model on one of its critical lines or, equivalently, a quantum chain of  $SU(2)_{N_f}$ spin-$\frac12$ anyons with ferromagnetic pair interaction favouring fusion in the $j=1$ channel \cite{BaRe89,GATL09}.
	
\subsection{$zH\gtrsim M_0$}
\begin{figure}[ht]
  \includegraphics[width=0.6\textwidth]{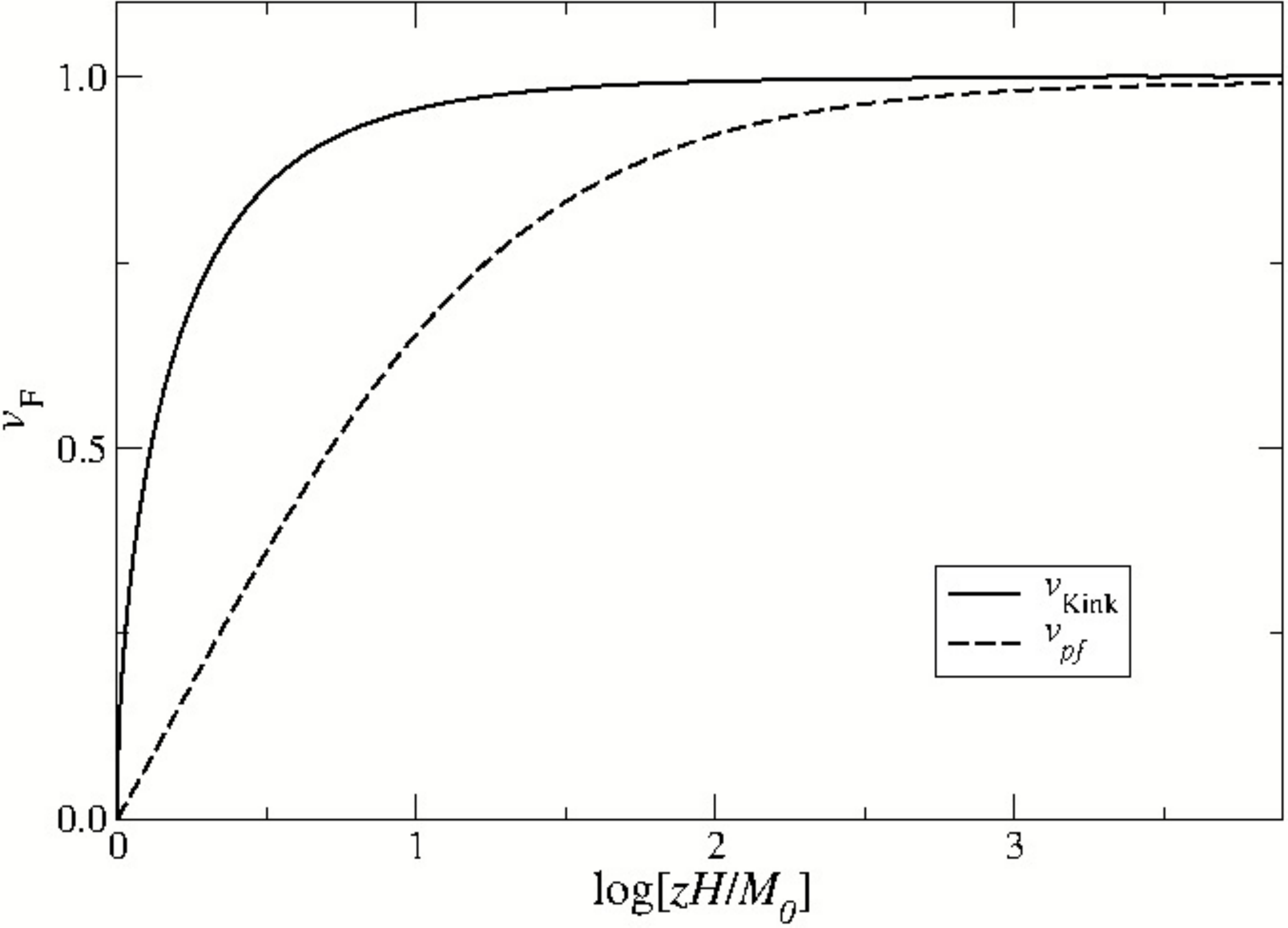}

    \caption{Fermi velocities of kinks and parafermion modes as a function of the magnetic field $zH>M_0$ for $p_0=2+1/3$ at zero temperature.  For large field, $H>H_\delta$, both Fermi velocities approach $1$ leading to the asymptotic result for the low temperature entropy (\ref{Ent_CFT}).
    \label{fig:FermiV}}
\end{figure}

\begin{figure}[th]
    \includegraphics[width=0.6\textwidth]{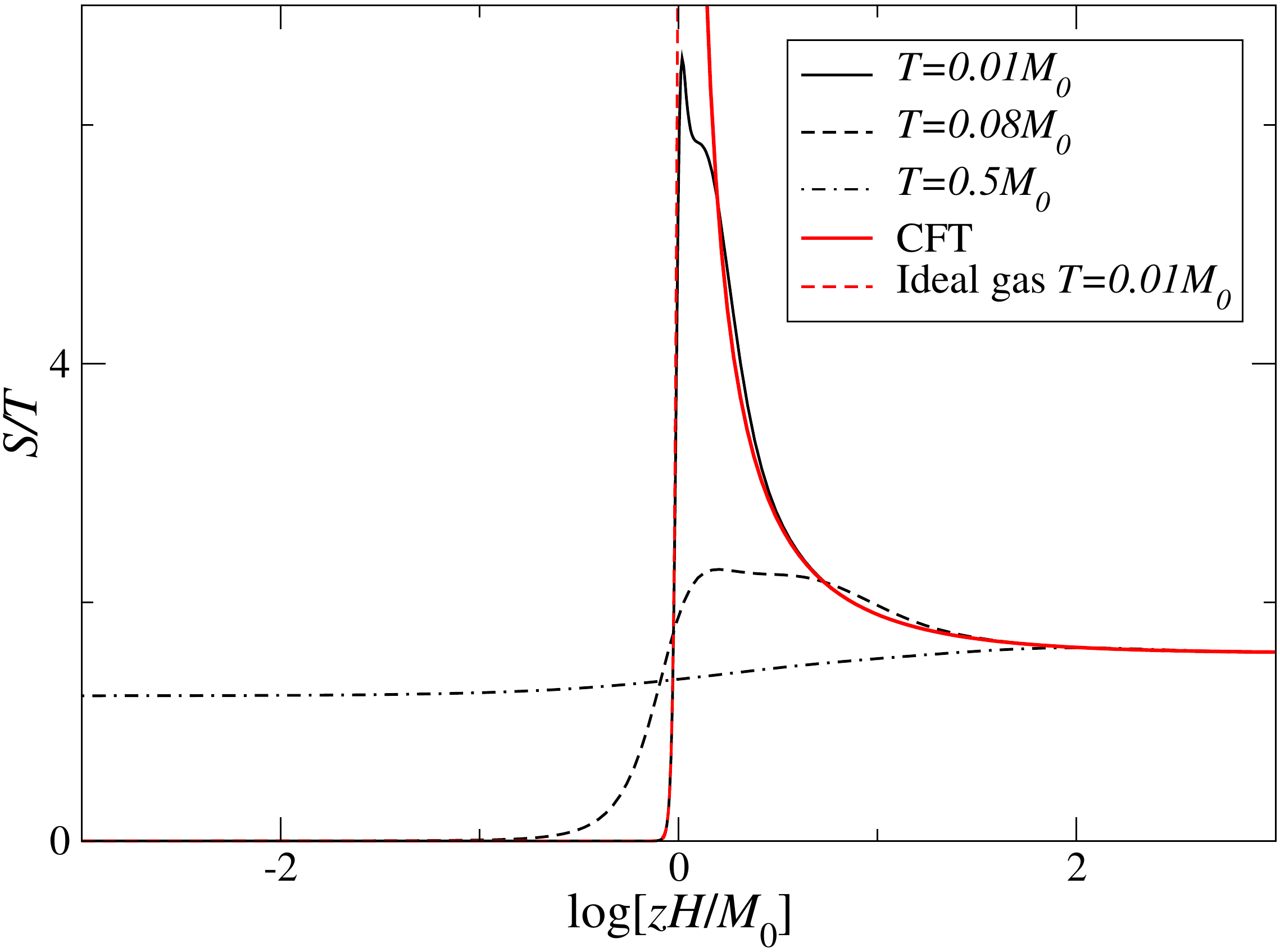}
    \includegraphics[width=0.6\textwidth]{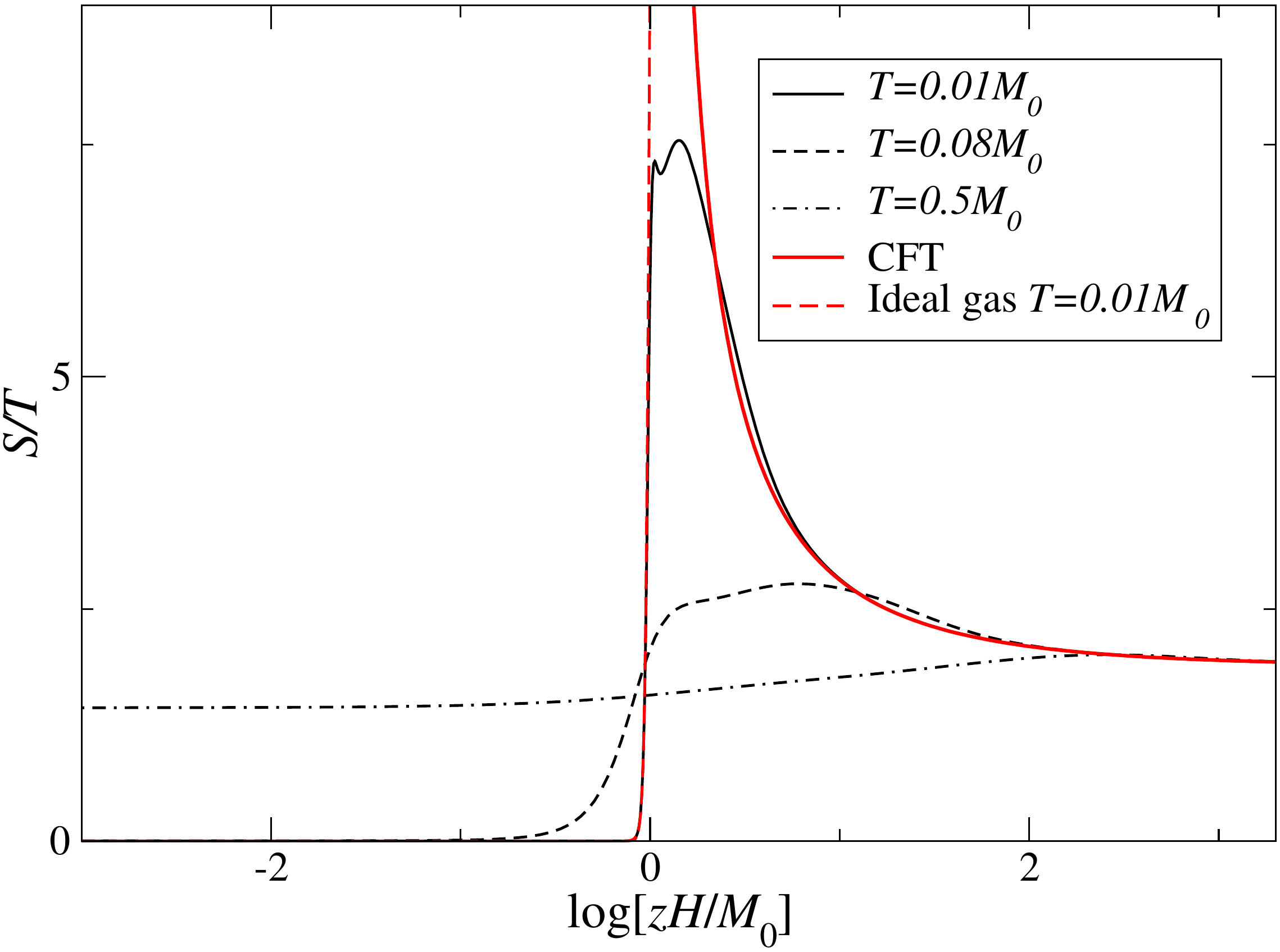}

  \caption{Entropy obtained from numerical solution of the TBA equations (\ref{dressedenergyinteq}) for $p_0=2+1/3$ (top), $p_0=3+1/3$ (bottom) as a function of the magnetic field $H$ for different temperatures.   For magnetic fields large compared to the kink mass, $zH\gg M_0$, the entropy approaches the expected analytical value (\ref{Ent_inter}) for a field theory with a free bosonic and a $Z_{N_f}$ parafermion sector propagating with velocities $v_{\text{Kink}}$ and $v_{pf}$, respectively (full red line).  For magnetic fields $zH<M_0$ and temperature $T\ll M_0$ the entropy is that of a dilute gas of non-interacting quasi-particles with degenerate internal degree of freedom due to the anyons (dashed red line).
    \label{fig:entropy}}
\end{figure}

To study the transition from the gas of free anyons to the condensate of kinks described by the CFT given above we have solved the TBA equations (\ref{dressedenergyinteq}) numerically by an iterative method for a given magnetic field $H$, temperature $T$ and anisotropy parameter $p_0$.
Using (\ref{freeenergy}) the entropy can be computed from the numerical data as
\begin{equation}
    \label{entropy_num}
    \begin{aligned}
        \mathcal{S} &= -\frac{\text{d}}{\text{d}T}\frac{F}{N}\\
        &= \sum_{j\notin\{j_2\}} \frac{M_j}{4}\int_{-\infty}^{\infty}\text{d}\lambda\, \cosh\left(\frac{\pi\lambda}{2}\right) \left(\log\left(1+e^{-\epsilon_j/T}\right)+\left(\frac{\epsilon_j}{T}-\frac{\text{d}}{\text{d}T}\epsilon_j\right)\left(1+e^{\epsilon_j/T}\right)^{-1} \right).
    \end{aligned}
\end{equation}
For magnetic fields large enough to suppress the contribution of breathers to the entropy, $zH \gg M_{0}$, the entropy is seen to be given by the term $j=j_0$ in (\ref{entropy_num}) alone and to converge to the expected analytical value (\ref{Ent_CFT}).  In the intermediate regime, $M_0<zH\lesssim zH_\delta$, and temperatures below the smallest of the breather masses $T\ll M_{N_f}$, the entropy deviates from this asymptotic expression: in this range of $H$ the auxiliary modes propagate with a  velocity (independent of $j_2$) differing from that of the kinks, $v_{\text{kink}}$, namely
\begin{equation}
    v_{pf} = -\left.\frac{ \partial_\lambda \epsilon_{j_2}(\lambda) }{2\pi\rho^h_{j_2}(\lambda)}\right|_{\lambda\to\infty}\,.
\end{equation}
As a consequence the bosonic (spinon) and parafermionic degrees of freedom separate and the low temperature entropy is
\begin{equation}
\label{Ent_inter}
    \mathcal{S} = \frac{\pi}{3} \left( \frac{1}{v_\text{kink}} + \frac{c(Z_{N_f})}{v_{pf}} 
    \right) T\,.
\end{equation}
Note that both Fermi velocities depend on the magnetic field and approach $1$ as $H\gtrsim H_\delta$, see Fig.~\ref{fig:FermiV}, giving the entropy (\ref{Ent_CFT}) of the WZNW model.
In Fig.~\ref{fig:entropy} the computed entropy (\ref{entropy_num}) is shown for various temperatures as a function of the magnetic field together with the $T\to0$ behaviour (\ref{Ent_inter}) expected from conformal field theory. Additionally, Fig.~\ref{fig:entropy} shows the decoupling into bosonic and parafermionic modes for $zH\gtrsim M_0$.\footnote{%
For low temperatures and magnetic fields $zH > M_0$ the contribution of kinks in (\ref{entropy_num}) develops a singularity which prevents reaching the asymptotic regime in the numerical analysis.}

\section{Summary and Conclusion}
We have studied the contribution of the spin degrees of freedom to the low temperature properties of a system of interacting fermions with spin and orbital degrees of freedom.  The excitations of this model are kinks connecting its topologically different ground states.  Localized on these kinks are modes with non-integer quantum dimension which have been identified as non-Abelian spin-$\frac12$ anyons satisfying $SU(2)_{N_f}$ fusion rules, among them Ising and Fibonacci anyons for $N_f=2$ and $3$, respectively \cite{GATL09,GATH13}.  By variation of an external magnetic field the density of kinks in the system can be controlled which allows to drive the system into a phase where kinks condense and the collective behaviour of the anyons can be studied once their interactions become important.

We summarize our findings for the contribution of the anyonic degrees of freedom to the low temperature thermodynamical properties of the system in Figure~\ref{fig:phasediag}:
\begin{figure}
    \includegraphics[width=0.8\textwidth]{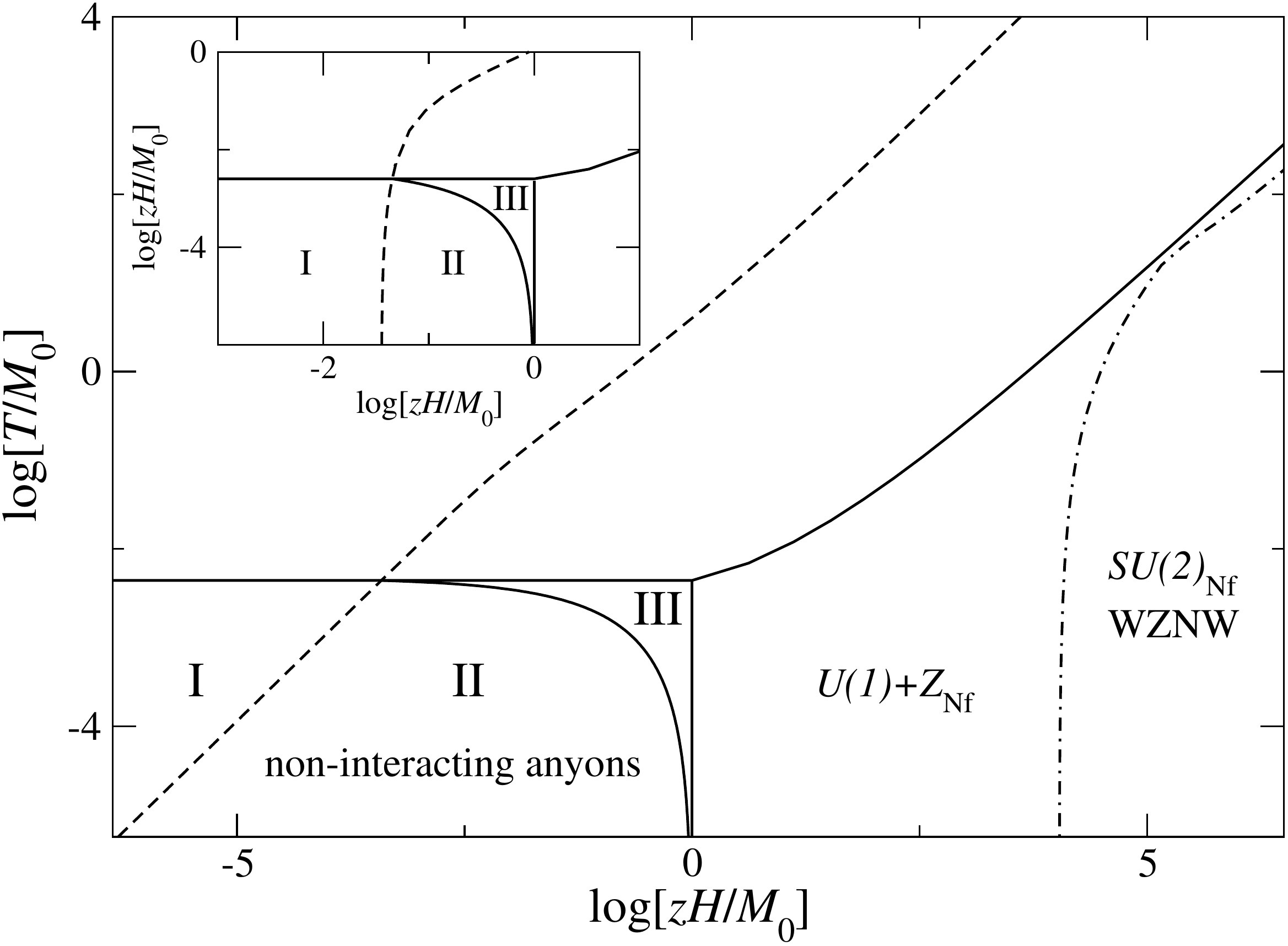}

  \caption{Contribution of the $SU(2)_{N_f}$ anyons to the low temperature properties of the model (\ref{NfModel}) for $\nu=3$ (inset for $\nu=4$): using the criteria described in the main text the parameter regions identified using analytical arguments for $T\to0$ before are located in the phase diagram (the actual location of the boundaries is based on numerical data for $N_f=2$).  For small magnetic fields a gas of non-interacting quasi-particles with the anyon as an internal zero energy degree of freedom bound to them is realized.  Here the dashed line indicates $\epsilon_{j_0}(0)=\epsilon_{N_f}(0)$, i.e. the location of the crossover between regions where the lowest energy breathers, $j_1=N_f$, (region I) or kinks (region II) dominate the free energy.  In region III the presence of thermally activated kinks with a small but finite density lifts the degeneracy of the zero modes.  As argued in Ref.~\cite{Tsve14a,JKLRT17} this results in the formation of a collective state of the anyons described by $Z_{N_f}$ parafermions. 
  For fields $zH>M_0$ the kinks condense and the low energy behaviour of the model is determined by the corresponding $U(1)$ bosonic mode and the parafermion collective modes of the non-Abelian $SU(2)_{N_f}$ spin-$\frac12$ anyons with ferromagnetic interaction.  For $zH\gg M_0$ the Fermi velocities of kinks and parafermions degenerate yielding a $SU(2)$ WZNW model at level $N_f$ for the effective description of the model.
  \label{fig:phasediag}}
\end{figure}

Signatures of $SU(2)_{N_f}$ non-Abelian anyons can be observed for temperatures small compared to the smallest breather mass, $T\ll M_{N_f}$, where the coupling to breathers $\{j_1\}$ can be neglected in the TBA equations (\ref{dressedenergyinteq}).  For magnetic fields $zH<M_0$ and temperatures small compared to the kink mass $T\ll M_0-zH$ the free energy (\ref{eq:Fideal}) of the system is that of an dilute gas of non-interacting kinks and breathers, both with an internal zero energy $SU(2)_{N_f}$ degree of freedom.  The main contribution comes from the lowest ($j_1=N_f$) breather for small magnetic fields while the kink becomes dominant once $\epsilon_{j_0}(0)<\epsilon_{N_f}(0)$.  As has been discussed above,  the degeneracy of the zero modes in the corresponding regions of the phase diagram (labelled I and II Figure~\ref{fig:phasediag}) can be identified with the quantum dimensions of $SU(2)_{N_f}$ anyons implying that a pair of these anyons with spin-$\frac12$ is bound to a breather while a single one is bound to a kink.

The kink gap closes as the magnetic fields is increased to $zH\lesssim M_0$.  Therefore the coupling between the kink and the auxiliary (anyon) modes in the TBA equations can no longer be neglected for temperatures $M_0-zH\lesssim T \ll M_{N_f}$ (region III in Figure~\ref{fig:phasediag}) where the distance between kinks becomes finite and higher order terms in their density have to be taken into account.  As a result the anyon modes begin to overlap and their degeneracy is lifted.  The resulting TBA equations resemble those of the critical RSOS models, whose low energy theory is a $Z_{N_f}$ parafermion CFT \cite{Tsve14a,JKLRT17}.  This is a first indication for the formation of a collective state by the anyons with ferromagnetic pair interaction which becomes manifest at magnetic fields $zH>M_0$: here the kinks acquire a finite density and the low energy modes of the system are particle-hole type excitations in the corresponding Fermi sea in addition to the collective excitations of the $SU(2)_{N_f}$ anyon degrees of freedom.  The resulting effective field theory description for low temperatures involves a product of a free boson and $Z_{N_f}$ parafermions describing the kinks and the collective modes of the anyons, respectively.  Increasing the magnetic field further beyond $H\simeq H_\delta$ the Fermi velocities of these sectors degenerate and the model becomes conformally invariant with the central charge being that of the $SU(2)$ WZNW model at level $N_f$.

\begin{acknowledgments}
We thank Andreas Kl\"umper and Alexei Tsvelik for discussions on this topic.  Funding for this work has been provided by the \emph{School for Contacts in Nanosystems}.  Additional support by the research unit \emph{Correlations in Integrable Quantum Many-Body Systems} (FOR2316) is gratefully acknowledged.
\end{acknowledgments}

\appendix
\section{Strings of the $XXZ$ Spin $S$ Model}
\label{app:stringclassification}
Following \cite{KiRe87} the admissible strings and parities of the $XXZ$ spin $S$ model are described by the numbers $p_i,b_i,y_i$ and $m_i$ with  $i\leq \ell-1$:
	\begin{alignat*}{4}
		&p_0=N_f+1/\nu \qquad\qquad &&p_1=1 \qquad\qquad &&b_i=[p_i/p_{i+1}]  &&p_{i+1}=p_{i-1}-b_{i-1}p_i\\
		&y_{-1}=0 \qquad\qquad &&y_0=1 \qquad\qquad &&y_1=b_0 \qquad\qquad &&y_{i+1}=y_{i-1}+b_iy_i \qquad\quad i\geq 0\\
		&m_0=0 \qquad\qquad &&m_1=b_0 \qquad\qquad &&m_{i+1}=m_i+b_i \qquad\quad && i\geq 0,
	\end{alignat*}
where $\ell$ is the length of the continued fraction expansion of the rational number $p_0$
	\begin{align*}
		&p_0=[b_0,b_1,b_2,\dots,b_{\ell-1}]=b_0+\frac{1}{b_1+\frac{1}{b_2+\dots}}\\
		&p_i/p_{i+1}=[b_i,b_{i+1},\dots,b_{\ell-1}].
	\end{align*}
In terms of these numbers the strings (\ref{string}) appearing in the solution of the Bethe equations for the XXZ spin-$\frac12$ chain have been classified by Takahashi and Suzuki \cite{TaSu72}: their lengths and parities are
\begin{equation}
	\label{lengthparity}
	\begin{aligned}
		&n_j=y_{i-1}+(j-m_i)y_i \quad\mathrm{for~} m_i\leq j <m_{i+1}\,,\\
		&v_j\equiv v_{n_j}=\exp\left(\mathrm{i}\left[\frac{n_j-1}{p_0}\right]\right) \qquad j\neq m_1\,,\\
		&
           v_{m_\ell}=(-1)^\ell\,,
	\end{aligned}
\end{equation}
for $1\leq j < m_\ell$ ($[x]$ denotes the integer part of $x$) and $n_{m_\ell}=y_{\ell-1}$, $v_{m_\ell}=(-1)^\ell$.
Provided that $2S+1$ appears in the sequence of Takahashi numbers $n_j$, i.e.\
	\begin{equation}
	    \label{condS}
		n_\sigma=2S+1\,, \quad m_r\leq \sigma <m_{r+1}\,,
	\end{equation}
the same classification holds for the integrable XXZ spin-$S$ chain related to the model (\ref{NfModel}) considered here \cite{KiRe87}.

For the definition of the kernels appearing in the Bethe ansatz integral equations we also need the sequences $q_j$ and $r(j)$:  for $m_i\leq j < m_{i+1}$ we define
	\begin{equation}
		q_j=(-1)^i[p_i-(j-m_i)p_{i+1}]\,, \quad r(j)=i\,, \quad
		q_{m_\ell}=-q_{m_\ell-1}\,.
	\end{equation}
Based on the physical properties of the corresponding excitations in the spin chain the strings are grouped into $j_0$-strings (kinks), $j_1$-strings (breathers) and $j_2$-strings (auxiliary functions) as  \cite{KiRe87}
\begin{equation}
	\begin{aligned}
		&\{j_0\} = \{m_i| i\leq r+1,\, i=r+1(\text{ mod } 2) \}\\
		&\{j_1\} = \{j|m_{i-1}\leq j<m_i,\, i\leq r+1,\, i=r+1(\text{ mod } 2)\}\\
		&\{j_2\}=\{\text{other strings}\}.
	\end{aligned}
\end{equation}
For the special choice of anisotropy parameter $p_0=N_f+1/\nu$ ($\nu>2$ and $\nu\in \mathbb{N}$) considered in this paper the sequences above become 
	\begin{alignat*}{4}
		&b_0=N_f, \quad &&b_1=\nu\\
		&p_0=N_f+1/\nu, \quad &&p_1=1, \quad &&p_2=1/\nu\,, \quad &&\\
		&m_0=0, \quad &&m_1=N_f, \quad &&m_2=N_f+\nu\,, \quad &&\\
		&y_{-1}=0, \quad &&y_0=1, \quad &&y_{1}=N_f\,, \qquad &&y_{2}=1+N_f\nu\,.
	\end{alignat*}
This yields the following admissible lengths and parities of the strings
	\begin{alignat}{4}
		&\text{aux. functions : }\quad &&n_{j_2}= {j_2}\,, \quad 
		    &&v_{j_2}=1\,, \qquad &&1\leq j_2<N_f\,,\nonumber\\
		&\text{breathers :}\quad &&n_{j_1}=(j_1-N_f)N_f+1\,, \quad
		    &&v_{j_1}=(-1)^{j_1+N_f+1}\,, \qquad &&N_f\leq j_1 <N_f+\nu\,,\\
		&\text{kinks :}\quad &&n_{j_0}=N_f\,, \quad &&v_{j_0}=1\,, \qquad &&j_0=N_f+\nu\,,\nonumber
	\end{alignat}
and $q_j$ numbers
	\begin{equation}
		q_{j_0}=1/\nu, \qquad q_{j_1}=\frac{1}{\nu}(j_1-N_f)-1,\qquad q_{j_2}=N_f+1/\nu-j_2.
	\end{equation}
Finally, since the Bethe equations (\ref{Betheeq}) can be related to those of the spin $S=N_f/2$ XXZ chain, the condition (\ref{condS}) is satisfied for $\sigma =N_f+1$, $r=1$.

\section{Definition of Kernels}
\label{AppendixB}
Following \cite{KiRe87} the Fourier transformed kernels $A_{jk}(\omega)$ of the $XXZ$ model are expressed in terms of the functions $a_j(\omega)$, $S_j(\omega)$ and $\zeta_j(\omega)$
\begin{equation}
	\begin{aligned}
		a_j(\omega)&=\frac{\sinh(q_j\omega)}{\sinh(p_0\omega)}\,, \qquad
		S_j(\omega)=\frac{1}{2\cosh(p_j\omega)}\,,\\
		\zeta_j(\omega)&=\cosh\left[\left(\left\{ \frac{n_j}{p_0}\right\}-\frac{1-(-1)^{r(j)}}{2}\right)p_0\omega\right]+ \sum_{\ell=1}^{n_j-1} \cosh\left[\left(\left\{\frac{n_j-\ell}{p_0} \right\}\left\{\frac{\ell}{p_0} \right\}\right)p_0\omega\right]		
	\end{aligned}
\end{equation}
where $\{x\}$ denotes the fractional part of $x$. This yields the Fourier transformed kernels $A_{jk}(\omega)$ and the Fourier transformed functions $a_{j,N_f}(\omega)$
\begin{equation}
	\label{xxzfunctions}
	\begin{aligned}
		&A_{kj}(\omega)=A_{jk}(\omega)=2a_k(\omega)\,\zeta_j(\omega)-\delta_{k,j_0}\delta_{j,j_0-1}\\
		&a_{j,N_f}(\omega)=A_{j,\sigma-1}(\omega)S_{r+1}(\omega)+2\cosh(q_\sigma\omega)\sum_{\ell=1}^{r}A_{j,m_\ell-1}(\omega)S_\ell(\omega)S_{\ell+1}(\omega).
	\end{aligned}
\end{equation}
The Fourier transformed kernels $B_{jk}(\omega)$ used for the integral equations (\ref{densities1}), (\ref{dressedeinteq}) are defined by
	\begin{alignat*}{3}
		B_{j_0j_0}&=\frac{1}{A_{j_0j_0}},\quad &&B_{j_0j_1}=\frac{A_{j_0j_1}}{A_{j_0j_0}},\quad &&B_{j_0j_2}=\frac{A_{j_0j_2}}{A_{j_0j_0}},\\
		B_{j_1k_1}&=\frac{A_{j_0j_1}A_{j_0k_1}}{A_{j_0j_0}}-A_{j_1k_1},\quad
		&&B_{j_1j_2}=\frac{A_{j_0j_1}A_{j_0j_2}}{A_{j_0j_0}}-A_{j_1j_2},\quad
		&&B_{j_2k_2}=A_{j_2k_2}-\frac{A_{j_0j_2}A_{j_0k_2}}{A_{j_0j_0}}
	\end{alignat*}
and the relations
\begin{equation*}
	\begin{aligned}
		B_{j_kj_0}&=-(-1)^{r(j_k)}B_{j_0,j_k}, \qquad k=1,2\,,\\
		B_{j_2j_1}&=-B_{j_1j_2}.
	\end{aligned}
\end{equation*}

\section{Scaling limit}
\label{app:scalinglim}
The scaling limit $g\ll 1$ of $\tilde{\rho}^{(0)}_{j_0}(\lambda)$ (\ref{relations1}) can be computed by first simplifying the expression in Fourier space
\begin{equation}
	\begin{aligned}
		\tilde{\rho}^{(0)}_{j_0}(\omega)&=\cos(\omega/g)\frac{a_{j_0,N_f}(\omega)}{A_{j_0j_0}(\omega)}\\
		&=\cos(\omega/g)\frac{\cosh((p_0-1)\omega)+\frac{\cosh (q_\sigma\omega)}{\cosh\omega}\sum_{l=0}^{N_f-2}\cosh((N_f-2l-1)\omega)}{2\sum_{\ell=0}^{N_f-1}\cosh((N_f-2\ell)\omega)\cosh(\omega/\nu)}\\
		&=\cos(\omega/g)\frac{1}{2\cosh\omega},
	\end{aligned}
\end{equation}
where we used the expressions (\ref{xxzfunctions}) and the sequences for $p_0=N_f+1/\nu$ given in Appendix~\ref{app:stringclassification}. After applying the inverse Fourier transformation to $\tilde{\rho}^{(0)}_{j_0}(\omega)$ we end up with
\begin{equation}
	\begin{aligned}
		\tilde{\rho}^{(0)}_{j_0}(\lambda)&=\frac{1}{8\cosh\left(\frac{\pi}{2}(\lambda+1/g)\right)}+\frac{1}{8\cosh\left(\frac{\pi}{2}(\lambda-1/g)\right)}\\
		&\stackrel{g\ll 1}{=}\frac{M_{0}}{4}\cosh\left(\frac{\pi \lambda}{2}\right), \qquad M_{0}=2e^{-\frac{\pi}{2g}}.
	\end{aligned}
\end{equation}
The scaling limit $g\ll 1$ for the other bare densities $\tilde{\rho}^{(0)}_j(\lambda)$ and energies $\tilde{\epsilon}^{(0)}_j(\lambda)$ can be performed in a similar way.
%
\bibliographystyle{apsrev4-1}

%

\end{document}